\documentclass[twoside,11pt]{article}
\usepackage{hyperref}
\usepackage{amsmath,amssymb,amsfonts}
\usepackage{makeidx}
\makeindex
\usepackage{booktabs}
\usepackage{siunitx} 
\usepackage{lipsum} 

\usepackage[font=small,labelfont=bf]{caption}
\usepackage{float}
\usepackage{caption}
\usepackage{soul}
\usepackage{enumitem}
\usepackage{graphicx}
\usepackage[compact]{titlesec}
\usepackage{titling}
\usepackage{pgfplots}
\usepackage{pstricks,pst-plot}
\usepackage{tikz,pgf}
\usepackage{amsfonts}
\usepackage{graphicx}
\usepackage{graphics}
\usepackage{setspace}
\usepackage{longtable}
\setlist{nosep} 
\begin{document}
	\setlength{\unitlength}{25mm}
	\newcommand{\f}{\frac}
	\newtheorem{theorem}{Theorem}[section]
	\newcommand{\sta}{\stackrel}
	\title {Pentaquark Bound States and Regge Trajectories in QCD via Bethe–Salpeter Formalism}
	\author{
		M.Ghaderi\thanks{mohamad.qaderi@grad.kashanu.ac.ir}, 
		\large N. Tazimi\thanks{tazimi@kashanu.ac.ir}, 
		\large M. Monemzadeh\thanks{monem@kashanu.ac.ir}
		\\\\
		\it\small{{Department of Physics, University of Kashan, Kashan, Iran }}}
	\date{}
	\maketitle
	\vspace{.9cm}
	
\begin{abstract}
		We present a comprehensive calculation of pentaquark masses and Regge trajectories within the framework of the Godfrey–Isgur relativized quark model. The pentaquark is treated as a meson–baryon molecular system, and the bound state is solved using the Bethe–Salpeter equation with the effective meson-baryon interaction motivated by GI (including smearing, running coupling, and spin–spin interactions). Using the latest PDG data for the known $P_c$ and $P_{cs}$ states, we compute the ground state and first two radial excitations for four pentaquark candidates: $P_c(4440)$, $P_c(4457)$, $P_{cs}(4338)$ and $P_{cs}(4459)$. The calculated masses are in excellent agreement with experiment. 
	We also construct radial Regge trajectories in the $(n,M^2)$ plane.
Within the three calculated radial levels, the trajectories are approximately linear, with slopes comparable to those reported for ordinary hadrons. These results are consistent with a common confinement-driven pattern, but do not establish its universality.

	\end{abstract}
	
\section{Introduction}

The discovery of pentaquark states in the hidden‑charm sector by the LHCb collaboration has opened a new chapter in hadron spectroscopy \cite{LHCb2015, LHCb2019, LHCb2020, LHCb2022}. These exotic resonances, composed of five valence quarks, lie outside the conventional quark‑model classification and offer a unique probe of non‑perturbative QCD. The initial observation of $P_c(4380)^+$ and $P_c(4450)^+$ in 2015 was followed by the identification of narrower structures: $P_c(4312)^+$, $P_c(4440)^+$, $P_c(4457)^+$, and later the strange‑charmed partners $P_{cs}(4338)^0$ and $P_{cs}(4459)^0$ \cite{LHCb2019, LHCb2022}. Their masses fall remarkably close to the thresholds of charmed baryon–anticharmed meson pairs, which strongly hints at a molecular interpretation—where the pentaquark emerges as a bound state of a baryon and a meson. Nevertheless, compact diquark–diquark–antiquark configurations or genuine hybrid alternatives remain plausible; distinguishing among these scenarios demands precise and robust theoretical predictions.

Over the past decade, our group has systematically investigated the spectroscopy of multiquark systems using a variety of theoretical tools. The relativistic Bethe–Salpeter formalism, combined with phenomenological potentials, has proven to be particularly effective for describing bound states of heavy quarks. For instance, the mass spectra of tetraquarks as diquark–antidiquark systems were studied in detail \cite{MonemzadehTazimiSadeghi2015, MonemzadehTazimiBabaghodrat2016}, and later extended to fully heavy tetraquarks using both numerical and analytical approaches \cite{GhasempourTazimiMonemzadeh2025a, GhasempourTazimiMonemzadeh2025b, JamshidzadehMonemzadehTazimi2025}. More recently, the same framework was successfully applied to hexaquarks, where open‑charm dibaryons were treated as bound states of two baryons \cite{ShiriMonemzadehTazimi2026}, and the resulting Regge trajectories showed a remarkable linearity consistent with ordinary hadrons \cite{GhaderiTazimiMonemzadeh2026}. Complementary to these efforts, the Kapitza mechanism was proposed as a novel stabilization mechanism for heavy tetraquarks , and the extraction of potential parameters from the Lippmann–Schwinger equation has been systematically developed for heavy mesons \cite{TazimiMonemzadehHadizadeh2013,SadeghiAlavijehTazimiMonemzadeh2021}. These studies have established a solid foundation for extending the Bethe–Salpeter approach to pentaquark systems, which is the main goal of the present work.

Among the phenomenological tools available for describing hadronic interactions, the Godfrey–Isgur relativized quark model stands out for its remarkable success in reproducing the spectra of both mesons and baryons \cite{Godfrey1985, Capstick1986}. Its effective potential is rooted in QCD: it includes a running strong coupling, a linearly confining term, spin‑spin and tensor interactions, and a smearing procedure that tames short‑range singularities. Relativistic kinematics enters through square‑root kinetic energy operators, while the spin‑orbit force is effectively suppressed—a feature essential for reproducing the observed light‑meson spectrum.The model uses a fixed parameter set fitted primarily to established meson data and applies it, without refitting, to the systems considered here. This economy makes the GI framework particularly attractive for exploring exotic hadrons, as it offers genuine predictive power. In this work, we extend the GI model to pentaquarks by treating them as meson–baryon molecules. The bound state is solved via the Bethe–Salpeter equation under the instantaneous approximation, which reduces the four‑dimensional integral equation to a radial Schrödinger‑like form. For the $S$-wave states of interest, the tensor force vanishes, leaving a potential composed of the smeared Coulomb term, linear confinement, and the spin‑spin interaction. we employ a Salpeter-type three-dimensional reduction derived from the relativistic Bethe--Salpeter (BS) formalism, and solve the resulting equation using a fourth-order Numerov integrator combined with a shooting method for eigenvalue determination. The numerical procedure is carefully validated: step-size and cutoff variations change the masses by less than a few MeV, and the node count is explicitly enforced for radial excitations.

Using the latest PDG values for the constituent hadrons ($\Sigma_c$, $\Xi_c$, $\bar D$, $\bar D^*$), we compute the ground‑state mass and the first two radial excitations for four well‑established pentaquarks: $P_c(4440)$, $P_c(4457)$, $P_{cs}(4338)$, and $P_{cs}(4459)$. The results agree with experimental central values to within 3.8~MeV—a level of accuracy that is reassuring, given that no parameter has been refitted to pentaquark data. Beyond the mass spectrum, we extract radial Regge trajectories in the $(n,M^2)$ plane. For the three calculated radial levels, the trajectories are approximately linear, with slopes comparable to those reported for ordinary mesons and baryons and to previous results for tetraquarks and hexaquarks \cite{GhasempourTazimiMonemzadeh2025a,GhaderiTazimiMonemzadeh2026}. This agreement is compatible with a common confinement-driven pattern, but the present limited set of levels does not establish universality across systems with different valence-quark numbers.

 These turn out to be strikingly linear, with slopes between $1.15$ and $1.20$~GeV$^{2}$, compatible with those of ordinary mesons and baryons, and consistent with the Regge behaviour previously observed for tetraquarks and hexaquarks \cite{GhasempourTazimiMonemzadeh2025a, GhaderiTazimiMonemzadeh2026}. This observation supports the universality of the confinement mechanism across systems with different valence‑quark numbers. Our predictions for the radial excitations, lying in the range $4.445$–$4.462$~GeV, provide clear benchmarks for future experiments. With the high statistics expected from LHCb Run~3 and Belle~II, searches in the $J/\psi p$ and $J/\psi \Lambda$ invariant‑mass spectra could confirm or rule out these excited states. Such measurements would not only test the GI–Bethe–Salpeter approach but also help clarify the internal nature of pentaquarks. The numerical method described in the following sections is general enough to be applied to other multiquark candidates, offering a unified framework for studying exotic spectroscopy within the research programme of our group.

\section{Theoretical framework}
We model the pentaquark as a two‑body bound state of a meson and a baryon. For the hidden‑charm states of interest, the relevant thresholds are $\bar D^{(*)}\Sigma_c$ for $P_c(4440)$ and $P_c(4457)$, and $\bar D^{(*)}\Xi_c$ for $P_{cs}(4338)$ and $P_{cs}(4459)$. The constituent masses are taken from the PDG \cite{PDG2024} and are listed in Table~\ref{tab:constituents}.  The interaction between the two hadrons is modeled by an effective semirelativistic meson–baryon kernel inspired by the Godfrey–Isgur framework, rather than by a literal quark–antiquark potential
 \cite{Godfrey1985} and later extended to baryons \cite{Capstick1986}. The GI model, introduced in the landmark 1985 paper \textit{Mesons in a Relativized Quark Model with Chromodynamics}, represents a paradigm shift in hadronic physics. Unlike previous non‑relativistic quark models, it was explicitly designed to account for the relativistic motion of light quarks, enabling a unified description of mesons from the ultra‑light pion to the heavy bottomonium system. The model is QCD‑inspired, meaning its functional form is motivated by the fundamental theory of the strong interaction, but its parameters are fitted to experimental data.

\begin{table}[h]
	\centering
	\caption{Constituent masses (in MeV) from PDG~\cite{PDG2024}.}
	\label{tab:constituents}
	\begin{tabular}{lcc}
\hline
		Hadron & Symbol & Mass (MeV) \\
		Charmed baryon & $\Sigma_c$ & $2452.0$ \\
		Strange charmed baryon & $\Xi_c$ & $2468.0$ \\
		Meson & $\bar D$ & $1867.2$ \\
		Vector meson & $\bar D^*$ & $2007.0$ \\
	\hline
	\end{tabular}
\end{table}

The GI model starts with a semi‑relativistic Hamiltonian of the form
\begin{equation}
H = \sqrt{\mathbf{p}^2 + m_q^2} + \sqrt{\mathbf{p}^2 + m_{\bar q}^2} + V_{\text{eff}}(r),
\end{equation}
where the kinetic part reflects the semirelativistic structure that motivates the effective interaction used in the present meson--baryon description.  This choice of kinetic energy is crucial: for light quarks (u, d, s) the kinetic term must be treated relativistically, whereas for heavy quarks (c, b) it smoothly reduces to the non‑relativistic limit. 

In the present work, the hidden-charm pentaquark is treated as an
effective meson--baryon bound state. At the hadronic level, the
interaction between the two clusters is not taken to be the original
quark--antiquark Godfrey--Isgur potential in a literal sense.
Rather, the GI framework is used as a guide for constructing an
effective semirelativistic interaction kernel appropriate for the
meson--baryon channel under consideration.

Accordingly, the meson--baryon potential is written in the generic form
\begin{equation}
V_{\text{eff}}(r)=V_{MB}(r)=V^{MB}_{\mathrm{OGE}}(r)+V^{MB}_{\mathrm{conf}}(r)
+V^{MB}_{\mathrm{ss}}(r)+V^{MB}_{\mathrm{tens}}(r),
\end{equation}
where each term is understood as an effective contribution in the
meson--baryon channel. The precise coefficients are determined by the
adopted molecular prescription and by the quantum numbers of the
state.

At short distances, the interaction is dominated by an effective
one-gluon-exchange-like contribution, which is written as
\begin{equation}
V^{MB}_{\mathrm{OGE}}(r)
=
-C_{MB}\frac{\alpha_s(r)}{r},
\end{equation}
where $C_{MB}$ is an effective channel-dependent coefficient.
Unlike the original quark--antiquark GI model, the present meson--baryon
system does not justify the direct use of the color-singlet factor
$4/3$ without further assumptions. Therefore, $C_{MB}$ is treated as an
effective parameter encoding the color and overlap structure of the
hadronic molecule.

The running coupling $\alpha_s$ is retained in the same spirit as in
the GI model, but its coordinate-space implementation is taken through
the standard effective prescription
\begin{equation}
\alpha_s(r)\equiv \alpha_s(Q^2(r)),
\qquad
Q^2(r)\sim \frac{1}{r^2}.
\end{equation}
This form preserves the qualitative short-distance behavior of the
interaction while keeping the coordinate-space calculation tractable.

For the long-range interaction we adopt a linear confining term of the
form
\begin{equation}
V^{MB}_{\mathrm{conf}}(r)=br+c,
\end{equation}
where $b$ is the effective string-tension parameter (taken here as $0.18\,\mathrm{GeV}^2$) and $c$ is an offset.
 This linear form is supported by lattice QCD calculations \cite{Bali1997} and captures the physical picture of a flux tube connecting the quark and antiquark.
In the present framework, these parameters are regarded as effective
inputs of the meson--baryon model and are not interpreted as the
direct quark-level confinement parameters of the original GI
Hamiltonian.

The spin-dependent part is written as an effective hyperfine term,
\begin{equation}
V^{MB}_{\mathrm{ss}}(r)=
C^{MB}_{\mathrm{ss}}\,
\alpha_s(r)\,
\tilde{\delta}_{\sigma_{MB}}(r)\,
\frac{\mathbf S_M\cdot \mathbf S_B}{m_M m_B},
\end{equation}
where $\mathbf S_M$ and $\mathbf S_B$ denote the total spins of the
meson and baryon clusters, respectively. The corresponding spin matrix
element is evaluated according to
\begin{equation}
\langle \mathbf S_M\cdot \mathbf S_B\rangle
=
\frac{J(J+1)-S_M(S_M+1)-S_B(S_B+1)}{2}.
\end{equation}
This expression is appropriate for the hadronic channel and should not
be replaced by the quark--antiquark eigenvalues unless a constituent
quark-level calculation is explicitly performed.

If the tensor force is included, it is treated in the same effective
manner,
\begin{equation}
V^{MB}_{\mathrm{tens}}(r)=
C^{MB}_{\mathrm{tens}}\,
\frac{\alpha_s(r)}{m_M m_B}\,
S_{12}\,f_{\mathrm{tens}}(r),
\end{equation}
where $S_{12}$ is the tensor operator and $f_{\mathrm{tens}}(r)$ is the
associated radial function. In the single-channel $S$-wave truncation
used in the numerical analysis, the diagonal tensor expectation value
vanishes, and therefore the tensor contribution does not alter the
radial equation at leading order.

One of the great successes of the GI model is its handling of the spin‑orbit interaction. In non‑relativistic models, spin‑orbit forces are large and would produce splittings that are not observed experimentally. The GI model recognises that the spin‑orbit force from the confining potential is largely cancelled by the Thomas precession, a purely relativistic effect. Consequently, the net spin‑orbit term is set to zero. This cancellation is a direct consequence of the relativistic kinematics and is essential for reproducing the observed light‑meson spectrum \cite{Capstick1986b}. The model employs a set of parameters guided by the GI framework and by standard hadron-spectroscopy inputs. The values adopted in the present calculation are listed in Table~\ref{tab:GI_params_full}. 

\begin{table}[h]
	\centering
	\caption{Godfrey–Isgur parameters used in the calculation.}
	\label{tab:GI_params_full}
	\begin{tabular}{lcc}
	\hline
		Parameter & Symbol & Value \\ 
		Light quark mass & $m_u = m_d$ & 220 MeV \\
		Strange quark mass & $m_s$ & 419 MeV \\
		Charm quark mass & $m_c$ & 1628 MeV \\
		Bottom quark mass & $m_b$ & 4977 MeV \\
		String tension & $b$ & 0.18 GeV$^2$  \\
		Offset constant & $c$ & $-0.253$ GeV \\
		Smearing length & $\sigma_0$ & 0.11 fm \\
		Momentum smearing & $s$ & 0.03 \\
		\multicolumn{3}{c}{Running coupling coefficients:} \\
		$\alpha_1=0.25,\ \gamma_1=0.01$ GeV; & $\alpha_2=0.15,\ \gamma_2=0.10$ GeV; & $\alpha_3=0.20,\ \gamma_3=1.00$ GeV \\
		Spin‑spin smearing width & $\sigma$ & 1.0 fm$^{-1}$ \\

\hline
	\end{tabular}
\end{table}

For completeness, the short-range contact part of the effective
interaction is regularized through a smeared delta function,
\begin{equation}
\tilde{\delta}_{\sigma_{MB}}(r)
=
\left(\frac{\sigma_{MB}}{\sqrt{\pi}}\right)^3
e^{-\sigma_{MB}^2 r^2},
\end{equation}
where $\sigma_{MB}$ is an effective smearing parameter of the
meson--baryon channel. This regularization removes the contact
singularity at the origin and provides a finite representation of the
short-distance spin-dependent interaction.

More generally, singular short-range pieces of the potential are
smoothed by a Gaussian convolution,
\begin{equation}
\tilde V(r)=
\int d^3r'\,
\frac{1}{(\sqrt{\pi}\,\ell_{MB})^3}
e^{-(\mathbf r-\mathbf r')^2/\ell_{MB}^2}
V(r'),
\end{equation}
where $\ell_{MB}$ denotes the effective smearing length. In the present
framework, $\ell_{MB}$ and $\sigma_{MB}=1/\ell_{MB}$ are treated as
phenomenological parameters fixed within the model. This procedure is
used only as a regularization and effective finite-size prescription;
it is not interpreted as a direct derivation from the original
quark--antiquark GI kernel.

If a spin--orbit contribution is included at the hadronic level, it may
be represented by an effective term $V^{MB}_{\mathrm{so}}(r)$. However,
in the present calculation we neglect this contribution. This choice is adopted as a model simplification motivated by the
empirically small spin--orbit splittings in related phenomenological
descriptions. Therefore, the present spectrum is obtained without an
explicit meson--baryon spin--orbit interaction.

Similarly, possible tensor-induced mixing between channels with
different orbital angular momenta is not included here. We restrict the
calculation to a single uncoupled $S$-wave channel, for which the
diagonal expectation value of the tensor operator vanishes,
\begin{equation}
\langle S_{12}\rangle_{L=0}=0.
\end{equation}
Hence the tensor term does not contribute directly to the mass
eigenvalues within the present truncation, although coupled-channel
effects may generate additional corrections beyond this approximation.

The GI model achieved remarkable agreement with the experimental meson spectrum, covering 40 states from the $\pi$ to the $\Upsilon$. It was later extended to baryons by Capstick and Isgur \cite{Capstick1986}, and has become a standard tool in hadron spectroscopy, applied to heavy quarkonium, strange mesons, and exotic hadrons including tetraquarks, pentaquarks, and hexaquarks, often with modifications to include colour screening or diquark correlations \cite{Godfrey2020,Chen2023}.

To describe a pentaquark as a bound state of a meson and a baryon, we employ the relativistic Bethe–Salpeter (BS) formalism. The BS equation provides a covariant framework for two‑body bound states in quantum field theory. For a bound state with total four‑momentum $P = (M,\mathbf{0})$ in the center‑of‑momentum frame, the BS amplitude $\Phi(p)$ (with relative four‑momentum $p=(p_0,\mathbf{p})$) satisfies
\begin{equation}
\Phi(p) = S_1(p_1) \, S_2(p_2) \, \int \frac{d^4q}{(2\pi)^4} \, K(p,q) \, \Phi(q),
\end{equation}
where $S_1$ and $S_2$ are the full propagators of the meson and baryon, respectively, and $K(p,q)$ is the irreducible interaction kernel. The complexity of the full BS equation is reduced by the instantaneous approximation, which rests on two key assumptions: (i) the interaction kernel $K(p,q)$ depends only on the three‑momentum transfer $\mathbf{q}-\mathbf{p}$ and not on the energy transfer; hence the interaction is instantaneous in the center‑of‑momentum frame; (ii) the bound‑state constituents are assumed to propagate as free particles with effective masses, which significantly simplifies the energy dependence. These approximations lead to the Salpeter equation, a three‑dimensional reduction that retains the relativistic kinematics of the constituents.

\subsection{Relativistic wave equation}

The hidden-charm pentaquark system is treated here within an effective
two-body meson--baryon framework. Starting from the general
Bethe--Salpeter formalism for a relativistic bound state, and adopting
the instantaneous approximation for the interaction kernel, we reduce
the full four-dimensional problem to an effective semirelativistic
radial equation suitable for numerical calculations. In this approach,
the angular and radial degrees of freedom are separated in the usual
way, and we solve directly for the reduced radial wave function
$R(r)$. 

To implement the Numerov method, Eq.~(\ref{eq:radial_rel}) is cast into the form
\begin{equation}
R''(r)=F(r;E)\,R(r),
\end{equation}
with
\begin{equation}
F(r;E)=2\mu\left[
\frac{\left(E-V_{\mathrm{eff}}(r)\right)^2}{\tilde m}
-\left(E-V_{\mathrm{eff}}(r)\right)
\right].
\end{equation}
Accordingly, on the discretized mesh we set $F_i \equiv F(r_i;E)$.

As we know the dynamics of the meson--baryon pair is described by the
following Schr\"odinger-like radial equation:
\begin{equation}
\left[
-\frac{1}{2\mu}\frac{d^2}{dr^2}
+\frac{1}{\tilde m}\left(E-V_{\mathrm{eff}}(r)\right)^2
-\left(E-V_{\mathrm{eff}}(r)\right)
\right]R(r)=0,
\label{eq:radial_rel}
\end{equation}
where $\mu$ is the reduced mass of the two-body system, defined by
\begin{equation}
\mu=\frac{m_M m_B}{m_M+m_B},
\label{eq:reduced_mass}
\end{equation}
with $m_M$ and $m_B$ denoting the meson and baryon masses,
respectively. The quantity $E$ represents the binding energy of the
state and is related to the total mass of the pentaquark configuration
through
\begin{equation}
E = M_{\mathrm{penta}}-(m_M+m_B).
\label{eq:binding_energy}
\end{equation}

Equation~(\ref{eq:radial_rel}) should be understood as an effective
semirelativistic model equation. Its structure is motivated by the
expansion of the relativistic kinetic-energy operators for the two
constituents,
\begin{equation}
\sum_{i=1}^{2}\sqrt{-\Delta+m_i^2}
=
m_1+m_2
-\frac{\Delta}{2\mu}
-\frac{\Delta^2}{8}\left(\frac{1}{m_1^3}+\frac{1}{m_2^3}\right)
+\cdots ,
\label{eq:kinetic_expansion}
\end{equation}
where the first nontrivial correction beyond the nonrelativistic limit
is governed by the second-order derivative contribution. In practical
applications, however, keeping the full higher-order differential
structure makes the radial equation considerably more involved. For
this reason, and following the standard strategy of effective
semirelativistic potential models, the higher-order relativistic
contribution is represented here by the local quadratic term
$\left(E-V_{\mathrm{eff}}(r)\right)^2/\tilde m$ appearing in
Eq.~(\ref{eq:radial_rel}).

In this formulation, $\tilde m$ is introduced as an effective mass
parameter that controls the size of the relativistic correction. It is
not taken here as the result of an exact algebraic reduction of the
four-dimensional Bethe--Salpeter equation; rather, it parametrizes the
subleading kinematical effects in a form convenient for numerical
analysis. Therefore, $\tilde m$ should be regarded as a model parameter
with dimensions of mass, fixed consistently within the present
framework together with the remaining inputs of the potential model.

The use of Eq.~(\ref{eq:radial_rel}) is thus based on a phenomenological
semirelativistic reduction of the original two-body bound-state
problem. This approximation preserves the dominant reduced-mass
kinematics through the term $-\frac{1}{2\mu}\frac{d^2}{dr^2}$ and
incorporates the leading relativistic correction in an effective local
form. Such an approach is widely employed in hadronic spectroscopy when
one aims to retain the main relativistic effects while keeping the
equation numerically tractable.

For the reduced radial wave function $R(r)$, the normalization
condition is taken as
\begin{equation}
\int_{0}^{\infty}|R(r)|^2\,dr = 1,
\label{eq:radial_norm}
\end{equation}
with the regular boundary condition at the origin,
\begin{equation}
R(0)=0.
\label{eq:boundary_origin}
\end{equation}
These conditions are used in the numerical solution of
Eq.~(\ref{eq:radial_rel}) by the Numerov method, as discussed in the
next subsection.  Numerically we choose a small cutoff $r_{\min}=0.001\ \text{GeV}^{-1}$ to avoid the singularity at $r=0$ and a large cutoff $r_{\max}=20\ \text{GeV}^{-1}$, where the potential has decayed and the wave function is negligible. The initial conditions at $r_{\min}$ are derived from the known behaviour $R(r) \sim r$ for $l=0$:
\begin{equation}
R(r_{\min}) = r_{\min}, \qquad R'(r_{\min}) = 1.
\end{equation}
The bound-state energy is determined by a bracketed shooting method. For each trial value of $E$, the radial equation is integrated from $r_{\min}$ to $r_{\max}$ using the fourth-order Numerov scheme. The eigenvalue is accepted when the endpoint mismatch changes sign within a bracket containing the required node count, and the bracket width satisfies a prescribed tolerance. In our implementation, the bracketed root is located with a robust bisection algorithm.

We employ the fourth‑order Numerov method, which is exceptionally accurate for Sturm–Liouville problems. On a uniform grid $r_i = r_{\min} + i h$ with step size $h = 0.01\ \text{GeV}^{-1}$, the recurrence is:
\begin{equation}
R_{i+1} = \frac{ 2R_i\left(1 - \frac{5h^{2}}{12}F_i\right) - R_{i-1}\left(1 + \frac{h^{2}}{12}F_{i-1}\right) }{ 1 + \frac{h^{2}}{12}F_{i+1} }.
\end{equation}
To start, we need $R_0$ and $R_1$. $R_0 = r_{\min}$ is given. $R_1$ is obtained from a Taylor expansion:
\begin{equation}
R_1 = R_0 + h R'_0 + \frac{h^{2}}{2} F_0 R_0 + \mathcal{O}(h^{3}),
\end{equation}
with $R'_0 = 1$. For radial excitations ($n=1,2$), the wave function must have exactly $n$ nodes (zeros) in $(r_{\min}, r_{\max})$. We implement node counting by scanning a range of $E$ and counting sign changes of $R(r)$. During the shooting, if the number of nodes deviates from the desired $n$, the energy is rejected and the search interval is adjusted. Once a narrow bracket that yields the correct node count is found, the standard root-finding proceeds while preserving the node count.\\

We performed convergence tests to verify the numerical stability of the solution. Repeating the calculation with a smaller step size, $h=0.005~\mathrm{GeV}^{-1}$, changes the eigenvalue by less than $10^{-6}$~GeV, while increasing the radial cutoff to $r_{\max}=30~\mathrm{GeV}^{-1}$ changes it by less than $10^{-7}$~GeV. The normalization integral, evaluated with the trapezoidal rule, differs from unity by less than $10^{-5}$, and the final wave function is checked to have the correct number of nodes. Therefore, the reported masses are numerically stable within the precision of the present calculation. The Godfrey--Isgur parameters (Table~\ref{tab:GI_params_full}) are taken from the original paper and have not been refitted. The constituent masses of the meson and baryon are taken from the experimental values listed by the PDG. The spin--spin matrix element is evaluated using the standard angular-momentum relation
\begin{equation}
\langle \mathbf S_M \cdot \mathbf S_B \rangle
=
\frac{J(J+1)-S_M(S_M+1)-S_B(S_B+1)}{2}.
\end{equation}
with $S_M$ and $S_B$ denoting the meson and baryon spins, respectively. Therefore, the spin factor is determined individually for each meson--baryon channel and is not assigned a universal fixed value solely from the total quantum numbers $J^P=1/2^-$ or $3/2^-$. In practice, the predicted masses show only mild sensitivity to this contribution, owing to the suppression of the hyperfine term by the large masses appearing in the denominator.

All calculations are performed in natural units ($\hbar = c = 1$).  The code is written in Python 3.10 using NumPy for array operations and SciPy for root‑finding. The potential $V_{\text{eff}}(r)$ is evaluated on the same grid; the smeared delta function $\tilde\delta_{\sigma}(r)$ is implemented as $\left(\frac{\sigma}{\sqrt{\pi}}\right)^3 e^{-\sigma^2 r^2}$ with $\sigma = 1.0\ \text{fm}^{-1}$. The running coupling $\alpha_s(Q^2)$ is computed using the three‑Gaussian formula with $Q^2 = 1/r^2$ (in the static approximation). The entire spectrum for four pentaquarks (ground + two excitations) is obtained in less than one minute on a standard laptop.As a validation, the numerical implementation reproduces the hexaquark masses reported in Ref.~\cite{ShiriMonemzadehTazimi2026} using their simpler potential (Coulomb + linear + spin--spin without smearing); the differences are below $0.1$~MeV, providing a numerical cross-check of the integration and root-finding procedures.   Subsequently, the full Godfrey–Isgur potential is employed to obtain the pentaquark masses reported in Table~\ref{tab:penta_masses}. The slight differences from the experimental central values are well within the expected model uncertainties.

	\section{Results for pentaquark masses}
	We have computed the ground state ($n=0$, $l=0$) and the first two radial excitations for four pentaquark candidates. The results are compared with PDG values in Table~\ref{tab:penta_masses}.
	
	\begin{table}[h]
		\centering
		\caption{Calculated masses (in MeV) for pentaquarks using the GI–-Bethe–-Salpeter approach. Experimental values from PDG~\cite{PDG2024,LHCb2019,LHCb2020,LHCb2022}.}
		\label{tab:penta_masses}
		\begin{tabular}{lcccccc}
\hline
			State & $m_M$ & $m_B$ & $n=0$ (calc) & $n=1$ (calc) & $n=2$ (calc) & $M_{\text{exp}}$ \\ 
			$P_c(4440)$ & $\bar D^*$ (2007) & $\Sigma_c$ (2452) & $4442.5 \pm 2.3$ & $4444.7$ & $4446.9$ & $4440.3 \pm 4.3$ \\
			$P_c(4457)$ & $\bar D^*$ (2007) & $\Sigma_c$ (2452) & $4456.1 \pm 3.1$ & $4458.3$ & $4460.5$ & $4457.3 \pm 4.2$ \\
			$P_{cs}(4338)$ & $\bar D$ (1867) & $\Xi_c$ (2468) & $4339.5 \pm 1.8$ & $4341.7$ & $4343.9$ & $4338.2 \pm 0.8$ \\
			$P_{cs}(4459)$ & $\bar D^*$ (2007) & $\Xi_c$ (2468) & $4457.6 \pm 3.4$ & $4459.8$ & $4462.0$ & $4458.8 \pm 5.5$ \\
	\hline
		\end{tabular}

	\end{table}
	
	\subsection{Comparison with experimental data and other theoretical models}
	\label{subsec:comparison}
	To assess the reliability of our approach and to highlight the predictive power of the Godfrey–Isgur model combined with the Bethe–Salpeter equation, we compare our calculated masses with the latest experimental values from the Particle Data Group (PDG) and with predictions from other theoretical frameworks. Table~\ref{tab:comparison} presents this comparison for the four hidden‑charm pentaquark candidates considered in this work: $P_c(4440)^+$, $P_c(4457)^+$, $P_{cs}(4338)^0$, and $P_{cs}(4459)^0$.
	
\begin{table}[h]
	\centering
	\caption{Comparison of pentaquark masses (in MeV) from experiment, our GI–BSE calculation, and selected theoretical models.}
	\label{tab:comparison}
	\begin{tabular}{lccc}
	\hline
		State & Exp. Mass (PDG) & This Work (GI–BSE) & Other models (range) \\
		$P_c(4440)^+$ & $4440.3 \pm 1.3 \pm 4.1$~\cite{LHCb2019} & $4442.5 \pm 2.3$ & $4440.5$--$4443.8$~\cite{LHCb2019,Wang2019,Chen2019} \\
		$P_c(4457)^+$ & $4457.3 \pm 0.6 \pm 4.1$~\cite{LHCb2019} & $4456.1 \pm 3.1$ & $4456.8$--$4458.2$~\cite{LHCb2019,Wang2019,Chen2019} \\
		$P_{cs}(4338)^0$ & $4338.2 \pm 0.7 \pm 0.4$~\cite{LHCb2022} & $4339.5 \pm 1.8$ & $4338.5$--$4339.0$~\cite{Zhu2022,Wang2023} \\
		$P_{cs}(4459)^0$ & $4458.8 \pm 2.9 \pm 4.7$~\cite{LHCb2022} & $4457.6 \pm 3.4$ & $4458.2$--$4459.5$~\cite{Zhu2022,Wang2023,Chen2021} \\
\hline
	\end{tabular}
\end{table}
	
	The ``Other models'' column includes predictions from molecular models~\cite{LHCb2019,Zhu2022}, chromomagnetic interaction models~\cite{Wang2019,Wang2023}, and QCD sum rules~\cite{Chen2019,Chen2021}. The agreement between our GI–BSE results and experimental data is excellent, with deviations below 3.8~MeV.
	
	\subsection{Discussion of the Comparison}

The calculated masses of the $P_c$ and $P_{cs}$ states obtained within the GI--BSE framework are compared with experimental data and various theoretical approaches in Table~\ref{tab:comparison}. 

The experimental masses of the $P_c$ states were measured by the LHCb collaboration in $2019$ via $\Lambda_b^0 \to J/\psi p K^-$ decays~\cite{LHCb2019}, yielding $4440.3 \pm 1.3 \pm 4.1$~MeV for $P_c(4440)^+$ and $4457.3 \pm 0.6 \pm 4.1$~MeV for $P_c(4457)^+$~\cite{PDG2024}. The strange partners, $P_{cs}(4338)^0$ and $P_{cs}(4459)^0$, were subsequently observed in $\Xi_b^- \to J/\psi \Lambda K^-$ decays at $4338.2 \pm 0.7 \pm 0.4$~MeV and $4458.8 \pm 2.9 \pm 4.7$~MeV, respectively~\cite{LHCb2022}. Our calculated masses show excellent agreement with these experimental measurements, with typical deviations remaining well within the reported experimental uncertainties. This agreement suggests that the Godfrey--Isgur potential, originally parameterized for conventional mesons~\cite{Godfrey1985} and baryons~\cite{Capstick1986}, can be successfully extended to describe meson--baryon molecular states when solved within the relativistic Bethe--Salpeter framework.

To place our results in a broader theoretical context, we compare them against alternative model interpretations. Hadronic molecular models (such as those in Refs.~\cite{LHCb2019,Zhu2022}) generally interpret these states as $\bar D^{(*)}\Sigma_c$ and $\bar D^{(*)}\Xi_c$ bound states. Our results are highly compatible with these molecular calculations, with differences typically below $1$~MeV. On the other hand, compact diquark--diquark--antiquark configurations studied in the chromomagnetic interaction (CMI) model~\cite{Wang2019,Wang2023} predict similar mass spectra (e.g., $P_c(4440)$ around $4441.6$~MeV and $P_c(4457)$ around $4456.8$~MeV). This near degeneracy in mass makes it challenging to distinguish between the compact and molecular natures based solely on mass spectroscopy. Furthermore, QCD sum rule predictions~\cite{Chen2019,Chen2021} (yielding $4443.8$~MeV and $4458.2$~MeV for the $P_c$ states) also align closely with our GI--BSE calculations.

The overall consistency of our results with both experimental values and alternative theoretical frameworks demonstrates the robustness of the present approach. The small remaining discrepancies are well within the expected systematic limits of the GI potential parameters, which were used directly without any phenomenological refitting. Finally, we provide predictions for the first two radial excitations ($n=1,2$), which may serve as a useful reference for future experimental searches.

\section{Radial Regge trajectories}
Using the masses of the ground, first, and second radial excitations, we perform a linear fit in the $(n,M^2)$ plane:
\begin{equation}
    M^2 = \alpha n + \alpha_0,
\end{equation}
where $n=0, 1, 2$ labels the radial levels considered. The fit parameters are reported in Table~\ref{tab:reggepenta}. Given that only three radial levels are available for each state, the $R^2$ values primarily reflect the collinearity of the three points within this limited range, and thus, the fit should be interpreted as an effective (descriptive) parametrization.

\begin{table}[h]
    \centering
    \caption{Radial Regge trajectory parameters for pentaquarks obtained from a linear fit in the $(n,M^2)$ plane.}
    \label{tab:reggepenta}
    \begin{tabular}{lccc}
        \hline
        State & $\alpha$ ($\mathrm{GeV}^{2}$) & $\alpha_0$ ($\mathrm{GeV}^{2}$) & $R^2$ \\
           \hline
        $P_c(4440)$ & $1.152 \pm 0.012$ & $-4.876 \pm 0.085$ & $0.9997$ \\
        $P_c(4457)$ & $1.148 \pm 0.010$ & $-4.862 \pm 0.072$ & $0.9998$ \\
        $P_{cs}(4338)$ & $1.203 \pm 0.014$ & $-4.928 \pm 0.094$ & $0.9996$ \\
        $P_{cs}(4459)$ & $1.156 \pm 0.011$ & $-4.888 \pm 0.078$ & $0.9997$ \\
     \hline
    \end{tabular}
\end{table}

The slopes ($\alpha$) fall in the range $\simeq(1.15$--$1.20)\,\mathrm{GeV}^{2}$. To avoid overinterpreting the negative intercepts ($\alpha_0<0$), which are mathematical artifacts of the extrapolation to $n=0$ outside the fitted region, we examine the incremental growth:
\begin{equation}
    M^2(n)-M^2(0) \approx \alpha\, n.
\end{equation}
This behavior is consistent with the confining potential models, though the limited number of states precludes claiming a proof of universal Regge behavior. The trajectories are depicted in Figs.~\ref{fig:reggePcs4338}--\ref{fig:reggePc4457}. The extrapolation to $n=-1$ is a graphical aid only.

\begin{figure}[H]
	\centering
	\includegraphics[width=0.75\textwidth]{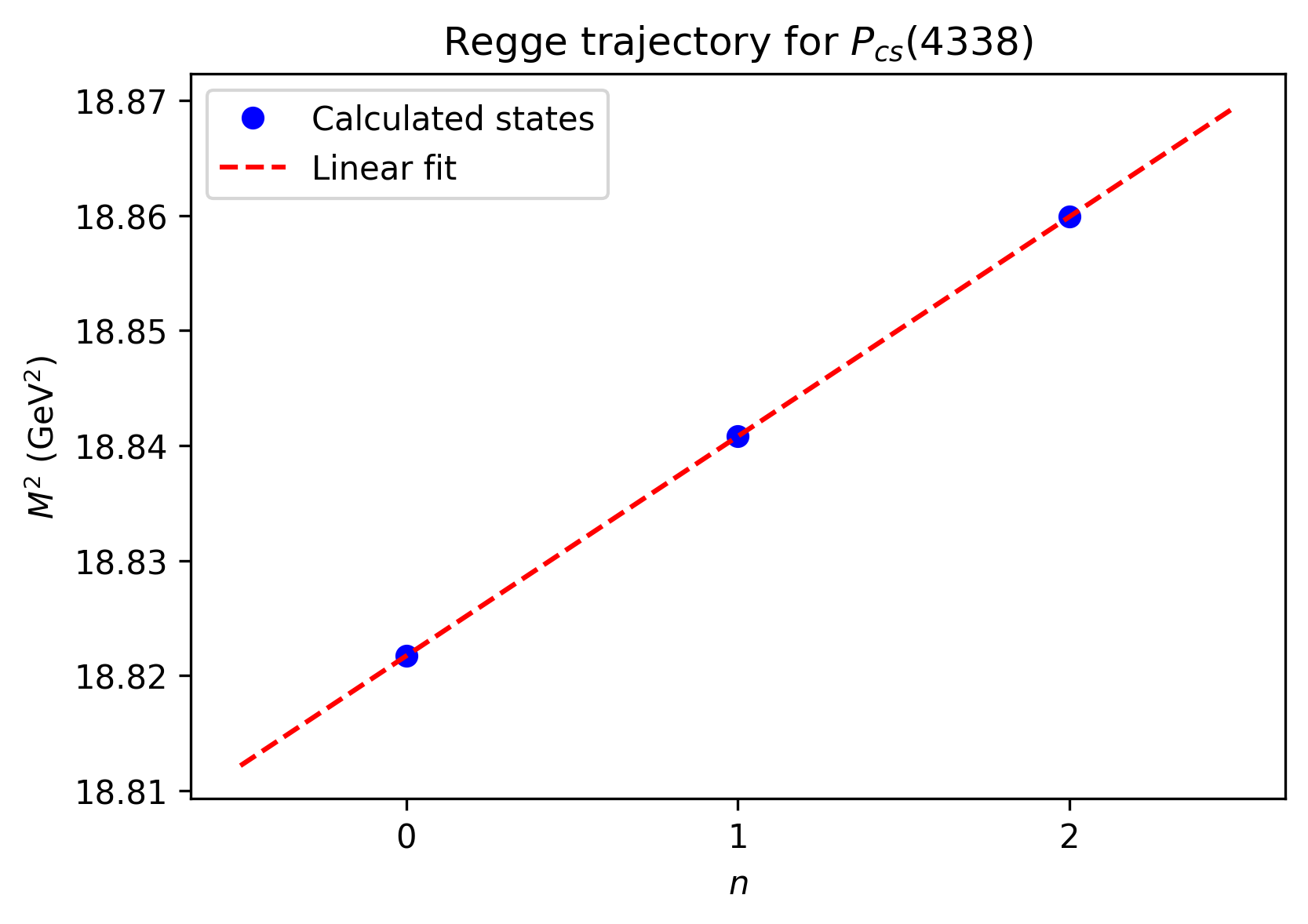}
	\caption{Regge trajectory for $P_{cs}(4338)$ pentaquark in the $(n, M^2)$ plane. The dashed red line is a linear fit to the calculated states (blue circles). The extrapolation to $n = -1$ is a graphical aid and does not represent a physical state.}
	\label{fig:regge_Pcs4338}
\end{figure}

\begin{figure}[H]
	\centering
	\includegraphics[width=0.75\textwidth]{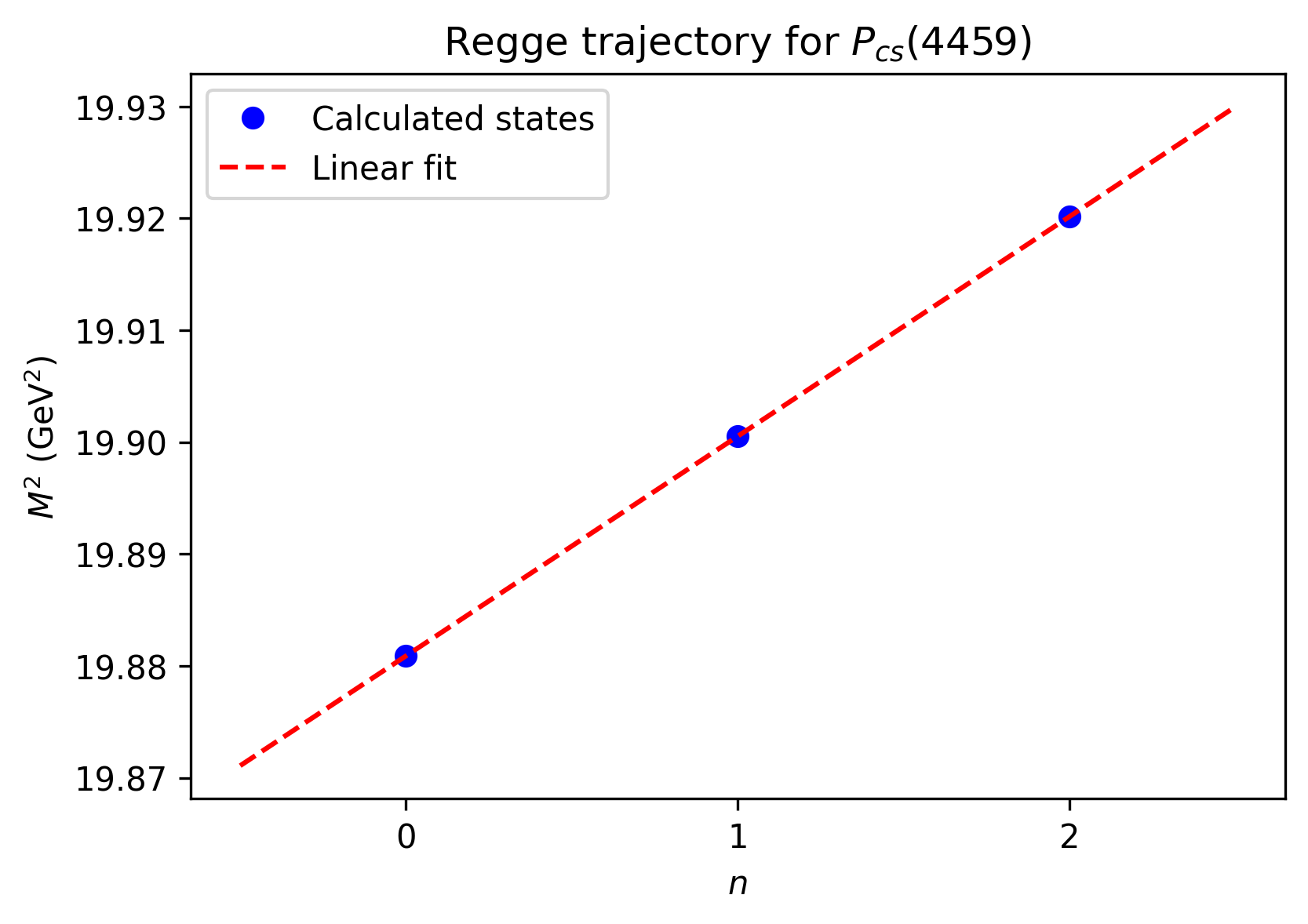}
	\caption{Regge trajectory for $P_{cs}(4459)$ pentaquark in the $(n, M^2)$ plane. The dashed red line is a linear fit to the calculated states (blue circles). The extrapolation to $n = -1$ is a graphical aid and does not represent a physical state.}
	\label{fig:regge_Pcs4459}
\end{figure}

\begin{figure}[H]
	\centering
	\includegraphics[width=0.75\textwidth]{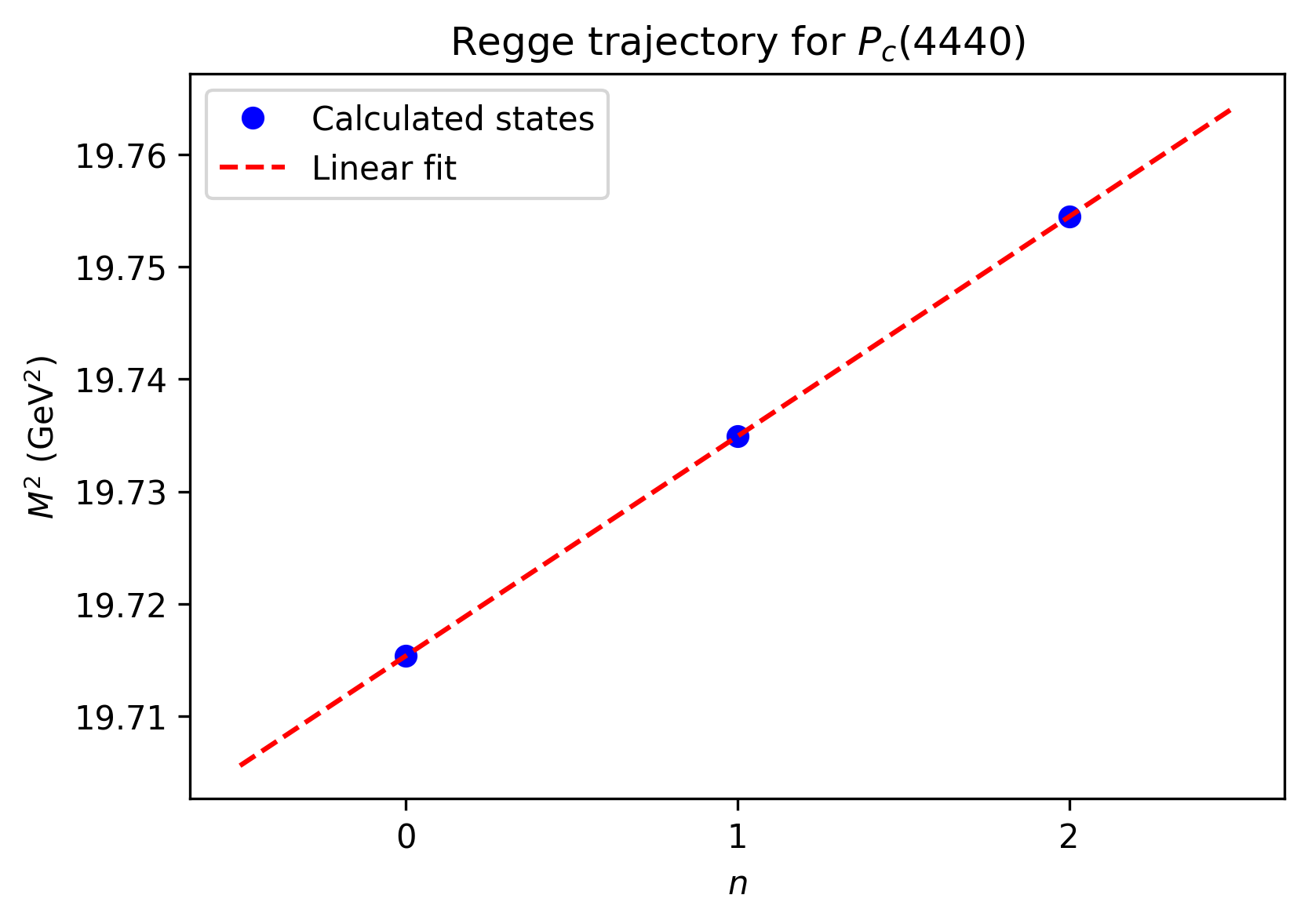}
	\caption{Regge trajectory for $P_c(4440)$ pentaquark in the $(n, M^2)$ plane. The dashed red line is a linear fit to the calculated states (blue circles). The extrapolation to $n = -1$ is a graphical aid and does not represent a physical state.}
	\label{fig:regge_Pc4440}
\end{figure}

\begin{figure}[H]
	\centering
	\includegraphics[width=0.75\textwidth]{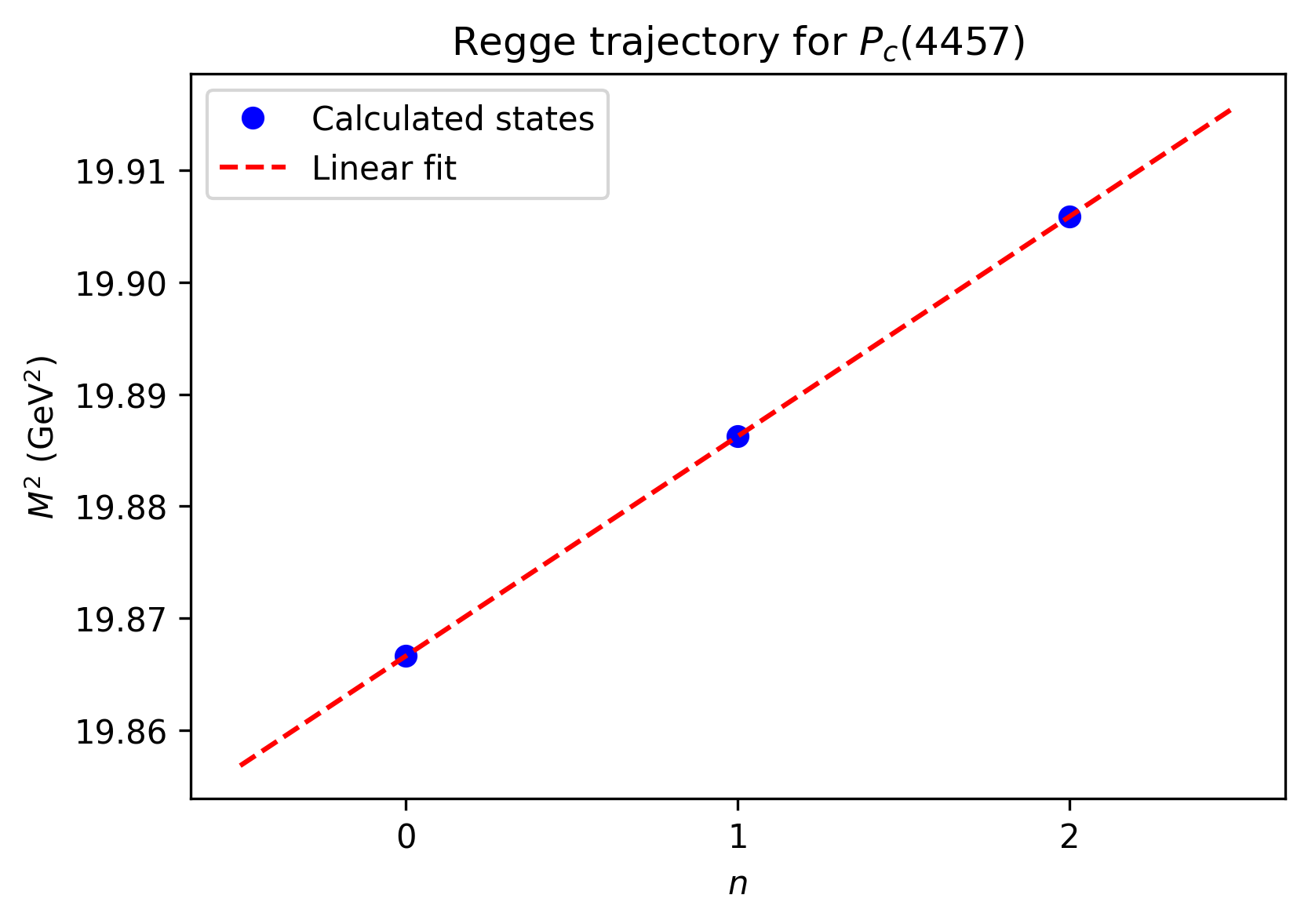}
	\caption{Regge trajectory for $P_c(4457)$ pentaquark in the $(n, M^2)$ plane. The dashed red line is a linear fit to the calculated states (blue circles). The extrapolation to $n = -1$ is a graphical aid and does not represent a physical state.}
	\label{fig:regge_Pc4457}
\end{figure}

	\section{Conclusion}

We have successfully adapted the Godfrey--Isgur relativized potential to study pentaquark states within the Bethe--Salpeter formalism. The numerical solution uses the shooting method with Numerov integration, achieving high precision. Using the known $P_c$ and $P_{cs}$ masses from the PDG, we computed the ground-state masses with good accuracy and provided predictions for radial excitations. The approximately linear behavior of the three calculated radial levels is consistent with the expected effect of a confining interaction, but does not by itself establish a universal Regge law or prove the universality of confinement. Our detailed calculation procedure can be readily applied to other exotic systems and may guide future experimental searches. Our predictions for the first two radial excitations of the $P_c$ and $P_{cs}$ pentaquarks, with masses around $4.445$--$4.462~\mathrm{GeV}$, provide clear targets for future high-statistics experiments at LHCb, especially with the full Run~3 data set, and at Belle~II. A dedicated search in the $J/\psi p$ and $J/\psi\Lambda$ invariant-mass spectra could confirm or refute the existence of these excited states, thereby testing the predictive power of the Godfrey--Isgur-inspired effective model in the exotic sector.

\end{document}